\documentclass{article}

\usepackage{times}
\usepackage{graphicx}
\usepackage{latexsym}

\usepackage{mathptmx}
\usepackage{multirow}
\usepackage{eepic,epic}
\usepackage{epsfig}
\usepackage{graphicx}
\usepackage{graphics}
\usepackage{color}
\usepackage{ifthen}
\usepackage{amssymb}
\usepackage[centertags]{amsmath}
\usepackage{pifont}
\usepackage{subfigure} 
\usepackage{paralist}
\usepackage[latin1]{inputenc} 
\usepackage{array,xspace,tikz}
\usepackage{tabularx,multirow}
\usepackage{booktabs}

\newenvironment{proofs}{\noindent{\bf Proof.}\hspace*{1em}}{\literalqed\bigskip}

\newcommand\qedblob{\mbox{\ding{113}}}
\def\literalqed{{\ \nolinebreak\hfill\mbox{\qedblob\quad}}}

\newcommand{\condition}{\,\mid \:}

\newcounter{alg}

\newcounter{case}

\newcounter{subcase}

\newcommand{\naturals}{\mathbb{N}}

\newtheorem{theorem}{Theorem}[section]

\newtheorem{lemma}[theorem]{Lemma}

\newcommand{\EP}[3]{
\begin{center}
{\small 
\begin{tabularx}{0.95\columnwidth}{ll}
\toprule
\multicolumn{2}{c}{\sc{#1}} \\
\midrule
{\bf Given:}& \parbox[t]{0.77\columnwidth}{#2\vspace*{1mm}} \\
{\bf Question:}& \parbox[t]{0.77\columnwidth}{#3\vspace*{.5mm}} \\ 
\bottomrule
\end{tabularx}
}
\end{center}
}

\newcommand{\np}{\ensuremath{\mathrm{NP}}}

\title{Taking the Final Step to a Full Dichotomy of the Possible Winner
  Problem in Pure Scoring Rules\footnote{This work was supported in part by
DFG grants RO-1202/11-1, RO-1202/12-1, and RO-1202/15-1, the European 
Science Foundation's EUROCORES program LogICCC, and the SFF grant
``Cooperative Normsetting'' of Heinrich-Heine-Universit{\"a}t
D{\"u}sseldorf.  A preliminary version appeared as a short
paper~\cite{bau-rot:c:dichotomy-possible-winner-scoring-rules} in
the proceedings of the \emph{19th European Conference on Artificial
Intelligence} (ECAI-2010).}}

\author{Dorothea Baumeister
\quad 
and
\quad 
J\"{o}rg Rothe
\\ Institut f\"ur Informatik
\\ Heinrich-Heine-Universit\"at D\"usseldorf
\\ 40225 D\"usseldorf, Germany
}
\begin{document}

\maketitle
\begin{abstract}
  The {\sc Possible Winner} problem asks, given an election where the voters'
  preferences over the candidates are specified only partially,
  whether a designated candidate can
become a winner
by suitably extending all the votes.
Betzler and
  Dorn~\cite{bet-dor:j:towards-dichotomy}
  proved a result that is
  only one step away from a full dichotomy of this problem for the
  important class of pure scoring rules in the case of unweighted
  votes and an unbounded number of candidates: {\sc Possible Winner}
  is $\np$-complete
  for all pure scoring rules except plurality, veto, and the scoring
  rule with vector $(2,1,\dots,1,0)$, but is solvable in polynomial
  time for plurality and veto.
  We take the final step to a full dichotomy by showing that
  {\sc Possible Winner} is
  $\np$-complete also for the scoring rule with vector
  $(2,1,\dots,1,0)$.
\end{abstract}

\section{Introduction}

The computational complexity of problems related to voting systems is
a field of intense study~(see, e.g., the surveys by Faliszewski et
al.~\cite{fal-hem-hem:j:cacm-survey,fal-pro:j:manipulation} and
Conitzer~\cite{con:j:making-decisions} and the bookchapters by
Faliszewski et al.~\cite{fal-hem-hem-rot:b:richer} and Baumeister et
al.~\cite{bau-erd-hem-hem-rot:b:computational-aspects-of-approval-voting}).
For many of the computational problems investigated, the voters are
commonly assumed to provide their preferences over the candidates via
complete linear orderings of all candidates.  However, this is 
not the case in many real-life settings: Some voters may have
preferences over 
some candidates only, or it may happen that new
candidates are introduced to an election after some voters have
already cast their votes.  As mentioned by
Chevaleyre et
al.~\cite{che-lan-mau-mon:c:possible-winners-new-candidates-scoring}
and Xia et
al.~\cite{xia-lan-mon:c:possible-winners-new-candidates-new-results},
such a situation
may occur, for example, when
a committee whose
members are to schedule their next meeting date by voting over a set
of proposed dates.  After some committee members have cast their votes
(and then have gone on vacation where they are unavailable via email
or phone), it turns out that some additional dates are possible, so
the remaining committee members have a larger set of alternatives to
choose from.  Since the meeting date has to be fixed before the
traveling committee members return from vacation, it makes sense to
ask whether the winning date can be determined via extending their
partial votes into complete linear ones by inserting the additional
alternatives.  Similar situations may also occur in large-scale
elections, where the computational aspects of the related problems
have more impact than for small-scale elections.

In light of such examples, it seems reasonable to assume only partial
preferences from the voters when defining computational problems
related to voting.  Konczak and Lang~\cite{kon-lan:c:incomplete-prefs}
were the first to study voting with partial preferences, and they
proposed the {\sc Possible Winner} problem that (for any given
election system) asks, given an election with only partial preferences
and a designated candidate~$c$, whether $c$ is a winner in
some extension of the partial votes to linear ones.  This problem was
studied later on by Xia and
Conitzer~\cite{xia-con:c:possible-necessary-winners}, Betzler and
Dorn~\cite{bet-dor:j:towards-dichotomy},
and Baumeister et
al.~\cite{bau-roo-rot:c:two-variants-of-possible-winner}, and closely
related problems have been introduced and investigated by Chevaleyre et
al.~\cite{che-lan-mau-mon:c:possible-winners-new-candidates-scoring},
Xia et
al.~\cite{xia-lan-mon:c:possible-winners-new-candidates-new-results},
and Baumeister et
al.~\cite{bau-roo-rot:c:two-variants-of-possible-winner}.  In
particular, Betzler and
Dorn~\cite{bet-dor:j:towards-dichotomy}
established a result that is only one step away from a full dichotomy
result of the {\sc Possible Winner} problem for the important class of
pure scoring rules.

Dichotomy results are particularly important, as they completely
settle the complexity of a whole class of related problems by
providing an easy-to-check condition that tells the hard cases apart
from the easily solvable cases.  The first dichotomy result in
computer science is due to Schaefer~\cite{sch:c:satisfiability} who
provided a simple criterion to distinguish the hard instances of the
satisfiability problem from the easily solvable ones.  Hemaspaandra
and Hemaspaandra~\cite{hem-hem:j:dichotomy-scoring} established the
first dichotomy result related to voting.  Their dichotomy result,
which distinguishes the hard instances
from the easy instances by the simple
criterion of ``\emph{diversity of dislike},'' concerns the
manipulation problem for the class of scoring-rule elections with
weighted votes.

In contrast, Betzler and Dorn's above-mentioned result that is just
one step away from a full dichotomy is concerned with the {\sc
  Possible Winner} problem for pure scoring rules with unweighted
votes and any number of
candidates~\cite{bet-dor:j:towards-dichotomy}.
In particular, they showed $\np$-completeness for all but three pure
scoring rules, namely plurality, veto, and the scoring rule with
scoring vector $(2,1,\dots,1,0)$.  For plurality and veto, they showed
that this problem is polynomial-time solvable, but the complexity of
{\sc Possible Winner} for the scoring rule with vector
$(2,1,\dots,1,0)$ was left open.  Taking the final step to a full
dichotomy result, we show that {\sc Possible Winner} is $\np$-complete
also for the scoring rule with vector $(2,1,\dots,1,0)$.

\section{Definitions and Notation}

An election $(C,V)$ is specified by a set $C=\{c_1,c_2,\dots,c_m\}$ of
candidates and a list $V=(v_1,v_2,\dots,v_n)$ of votes over~$C$.  In the
most common model of representing preferences, each such vote is a
linear order\footnote{
Formally, a \emph{linear order $L$ on $C$} is a binary relation on~$C$
that is
(i)~\emph{total} (i.e., for any two distinct $c,d \in C$, either
$c\,L\,d$ or $d\,L\,c$);
(ii)~\emph{transitive} (i.e., for all $c, d, e \in C$, if $c\,L\,d$
and $d\,L\,e$ then $c\,L\,e$); and
(iii)~\emph{asymmetric} (i.e., for all $c,d \in C$, if $c\,L\,d$ then
$d\,L\,c$ does not hold).
Note that asymmetry of $L$ implies \emph{irreflexivity of~$L$}
(i.e., for no $c \in C$ does $c\,L\,c$ hold).}
of the form $c_{i_1} > c_{i_2} > \dots > c_{i_m}$ where
$\{i_1,i_2, \dots, i_m\} = \{1, 2, \dots , m\}$, and $c_{i_k} >
c_{i_{\ell}}$ means that candidate $c_{i_k}$ is (strictly) preferred to
candidate $c_{i_{\ell}}$.  A voting system is a rule
to determine 
the winners of an election
Scoring rules (a.k.a.\ scoring protocols) are an important class of
voting systems.  Every scoring rule 
for $m$ candidates is specified
by a scoring vector $\vec{\alpha} =
(\alpha_1,\alpha_2,\dots,\alpha_m)$ with
$\alpha_1 \geq \alpha_2 \geq \dots \geq \alpha_m$,
where each $\alpha_j$ is a nonnegative integer.  For an election
$(C,V)$, each voter $v \in V$ gives $\alpha_j$ points to the candidate
ranked at the $j$th position in his or her vote.
Summing up all points a candidate
$c \in C$ receives from all votes in~$V$, we obtain
$score_{(C,V)}(c)$, $c$'s score in $(C,V)$.  Whoever has the highest
score wins the election.  If there is only one such candidate, he or
she is the unique winner.
Betz\-ler and Dorn~\cite{bet-dor:j:towards-dichotomy}
focus on
so-called pure scoring rules. A scoring rule is
\emph{pure} if
for each  
$m \geq 2$, the scoring vector for $m$ candidates can be obtained from
the scoring vector for $m-1$ candidates by inserting one additional
score value at any position subject to satisfying
$\alpha_1 \geq \alpha_2 \geq \dots \geq \alpha_m$.
We will study
only the
pure scoring rule that for $m\geq 2$
candidates is defined by the scoring vector $(2,1,\dots,1,0)$: In each
vote the
first candidate gets two points, the last candidate gets zero points,
and the $m-2$ other candidates get one point each.
We thus distinguish between the first, a middle, and the last position in
any vote.

The {\sc Possible Winner} problem is defined for
partial rather than
linear votes. For a set $C$ of candidates, a \emph{partial vote over
  $C$} is a transitive, asymmetric (though not necessarily total)
binary relation on~$C$.
For any two candidates $c$ and $d$ in a partial vote, we
write $c \succ d$ if $c$ is (strictly) preferred to~$d$.
For any two sets $A,B \subseteq C$ of candidates, we write $A \succ B$
to mean that each candidate $a \in A$ is preferred to each candidate
$b \in B$, i.e., $a \succ b$ for all $a \in A$ and $b \in B$.
As a shorthand, we write $a \succ B$ for $\{a\} \succ B$ and 
we write $A \succ b$ for $A \succ \{b\}$.

A linear vote $v'$ over $C$ \emph{extends} a partial vote $v$ over $C$
if $v \subseteq v'$, i.e., for all $c, d \in C$, if $c \succ d$ in $v$
then $c > d$ in~$v'$.  A list $V' = (v_1',v_2',\dots,v_n')$ of linear
votes over $C$ is an \emph{extension} of a list $V = (v_1,v_2,\dots,v_n)$ of
partial votes over $C$ if for each~$i$, $1 \leq i \leq n$, $v_i' \in V'$
extends $v_i \in V$.

Given a voting system $\mathcal{E}$,
Konczak and Lang~\cite{kon-lan:c:incomplete-prefs} define
the following problem:

\EP{$\mathcal{E}$-Possible Winner}
{A set $C$ of candidates, a list $V$ of partial votes over~$C$, and a
  designated candidate $c \in C$.}
{Is there an extension $V'$ of $V$ to linear votes over~$C$ such that
  $c$ is a winner of election $(C,V')$ under voting
  system~$\mathcal{E}$?}

This defines the problem in the nonunique-winner case; for
its unique-winner variant, simply replace ``a winner'' by ``the unique
winner.''
We focus on the nonunique-winner case here, but mention that the
unique-winner case can be handled analogously as described by Betzler
and
Dorn~\cite{bet-dor:j:towards-dichotomy}.
We may drop the prefix ``$\mathcal{E}$-'' and simply write
{\sc Possible Winner} when the specific voting system used is either
clear from the context or not relevant in the corresponding context.

\section{The Final Step to a Full Dichotomy Result}

Theorem~\ref{thm:321-NP} below shows that {\sc Possible Winner} for 
the scoring rule with vector $(2,1,\dots,1,0)$ is $\np$-hard.
Our proof of this theorem uses the notion of maximum partial score
defined by Betzler and
Dorn~\cite{bet-dor:j:towards-dichotomy}.
Fix any
scoring rule.  
Let $C$ be a set of candidates, $c \in C$ a candidate we want to make
win the election, and let $V=V^{\ell} \cup
V^p$ be a list of votes over~$C$, where $V^{\ell}$ contains only
linear votes and $V^p$ contains partial (i.e., incomplete) votes such that $c$'s
score is fixed, i.e., the exact number of points $c$ receives from any
$v \in V^p$ is known, no matter to which linear vote $v$ is extended.
For each $d \in C - \{c\}$, define
the \emph{maximum partial score of~$d$ with respect to~$c$} (denoted
by $s_p^{\max}(d,c)$) to be the maximum number of points that $d$ may
get from (extending to linear votes) the partial votes in $V^p$
without defeating $c$
in $(C,V')$ for any extension $V'$ of $V$ to linear votes.
Since the score of $c$ is the same in any extension $V'$
of $V$ to linear votes, it holds that
\begin{eqnarray*}
s_p^{\max}(d,c) & = & score_{(C,V')}(c)-score_{(C,V^{\ell})}(d).
\end{eqnarray*}
The following lemma will be useful for our proof of
Theorem~\ref{thm:321-NP}.

\begin{lemma}[Betzler and Dorn~\cite{bet-dor:j:towards-dichotomy}]
\label{lem:maximum-partial-scores}
  Let $\vec{\alpha} = (\alpha_1,\alpha_2,\dots,\alpha_m)$ be any
  scoring rule, let $C$ be a set of $m \geq 2$ candidates with
  designated candidate $c \in C$, 
  let $V^p$ be a list of partial
  votes in which the
  score of $c$ is fixed, 
and let $s_p^{\max}(c',c)$ be
the maximum partial score with respect to $c$ for all $c' \in C - \{c\}$.
  Suppose that the following two properties hold:
\begin{enumerate}
\item There is a candidate $d \in C - \{c\}$ such that $s_p^{\max}(d,c)
  \geq \alpha_1 |V^p|$.
\item For each $c' \in C - \{c\}$, the maximum partial score of $c'$
  with respect to $c$ can be written as a linear combination of the
  score values, $s_p^{\max}(c',c)=\sum_{j=1}^m n_j \alpha_j$,
 with $m=|C|$, $n_j \in \naturals$,
and $\sum_{j=1}^m n_j \leq |V^p|$.
\end{enumerate}

Then a list $V^{\ell}$ of linear votes can be constructed in
polynomial time such that for all $c' \in C - \{c\}$,
$score_{(C,V^{\ell})}(c')
 =
 score_{(C,V')}(c) - s_p^{\max}(c',c)$,
 where $V'$ is an extension of $V^p$ to linear votes.
\end{lemma}

\begin{theorem}
  \label{thm:321-NP} 
  {\sc Possible Winner} (both in the nonunique-winner case and in the
  unique-winner case) is $\np$-complete for the pure scoring rule with
  scoring vector $(2,1,\dots,1,0)$.
\end{theorem}

\begin{proofs}
Membership in $\np$ is obvious.  Our $\np$-hardness proof uses a
reduction from the $\np$-complete {\sc{}Hitting Set} problem (see, e.g.,
\cite{gar-joh:b:int}), which is defined as
follows:
\EP{Hitting Set}
{A finite set $X$, a collection $\mathcal{S} = \{S_1,\dots, S_n\}$ of
  nonempty subsets of $X$ (i.e., $\emptyset \neq S_i \subseteq X$ for
  each~$i$, $1 \leq i \leq n$), and a positive integer~$k$.}
{Is there a subset $X' \subseteq X$ with $|X'|\leq k$ such that $X'$
  contains at least one element from each subset in~$\mathcal{S}$?}

  Let $(X,\mathcal{S},k)$ be a given {\sc{}Hitting Set} instance with
  $X=\{e_1,e_2,\ldots,e_m\}$ and $\mathcal{S} = \{S_1, S_2, \ldots, S_n\}$.
  From $(X,\mathcal{S},k)$ we construct a {\sc Possible Winner} instance
  with candidate set
\[
C = \{c,h\}
    \cup \{x_i, x_i^1, x_i^2, \ldots, x_i^n,
           y_i^1, y_i^2, \ldots, y_i^n,
           z_i^1, z_i^2, \ldots, z_i^n
  \condition 1 \leq i \leq m\}
\]
  and designated candidate~$c$.  The list of votes $V = V^{\ell} \cup V^p$
  consists of a list $V^{\ell}$ of linear votes and a list $V^p$ of
  partial votes.  $V^p = V_1^p \cup V_2^p \cup V_3^p$ consists of three
  sublists:
\begin{enumerate}
\item $V_1^p$ contains $k$ votes of the form
$h\succ C - \{h,x_1,x_2,\ldots,x_m\} \succ \{x_1,x_2,\ldots,x_m\}$.
\item $V_2^p$ contains the following $2n+1$ votes for each~$i$, $1\leq
  i \leq m$:
\[
\begin{array}{rll}
  v_i:   & h \succ C - \{h,x_i,y_i^1\} \succ \{x_i,y_i^1\}, \\
  v_i^j: & y_i^j\succ C - \{y_i^j,z_i^j,h\} \succ h & \text{for } 1 \leq j
  \leq n, \\
  w_i^j: & x_i^j \succ C - \{x_i^j, y_i^{j+1},z_i^j\} \succ
  y_i^{j+1}& \text{for } 1 \leq j \leq n-1,\\
  w_i^n: & x_i^n \succ C - \{x_i^n,z_i^n,h\} \succ h.
\end{array}
\]
\item $V_3^p$ contains the vote
$T_j \succ C - \{T_j,h\} \succ h$ for each~$j$, $1 \leq j \leq n$,
where
\[
T_j = \{x_i^j \condition e_i \in S_j\}.
\]
\end{enumerate}

For each $i$, $1\leq i \leq m$, and~$j$, $1 \leq j \leq n$, the
maximum partial scores with respect to $c$ are set as follows:
 \begin{eqnarray*}
s_p^{\max}(x_i,c)     &  =  & |V^p|-1\\
s_p^{\max}(x_i^j,c)   &  =  & |V^p|+1\\
s_p^{\max}(y_i^j,c)   &  =  & s_p^{\max}(z_i^j) ~=~ |V^p|\\
s_p^{\max}(h,c)       &\geq & 2 |V^p|.
 \end{eqnarray*}
 This means that each $x_i$ must take at least one last position,
 which is possible in the votes from $V_1^p$ and the votes $v_i$,
 $1\leq i \leq m$, from $V_2^p$. Since the candidates $x_i^j$ can
 never take a last position, they may take at most one first position.
 For $y_i^j$ and $z_i^j$, the maximum partial scores with respect to $c$ are
 set such that for each first position they take, they must also take at
least one last position.
Finally, $h$ 
can never beat~$c$.
By Lemma~\ref{lem:maximum-partial-scores}, we can construct a list of
votes $V^{\ell}$ such 
that all candidates other than $c$ can get only their maximum partial
scores with respect to $c$ in
the partial votes.

 We claim that
 $(X,\mathcal{S},k)$ is a yes-instance of {\sc Hitting Set}
if and only if $c$
is a possible winner in $(C,V)$,
  using the scoring rule with vector
 $(2,1,\dots,1,0)$.

From left to right, suppose there exists a hitting set $X'
\subseteq X$ with $|X'| \leq k$ for~$\mathcal{S}$. The partial votes in
$V^p$ can then be extended 
to linear votes such that $c$ wins the
election as follows:
\[
\begin{array}{@{}l@{\hspace*{2mm}}r@{\hspace*{2mm}}l|@{\hspace*{2mm}}l@{\hspace*{2mm}}l@{}}
\hline
  & & e_i \in X' & e_i \not \in X'
\\ \cline{3-4}
  V_1^p: & & h > \dots > x_i
\\ \cline{3-4}
  V_2^p: & v_i: & h > \dots > x_i > y_i^1 & h > \dots > y_i^1 > x_i \\
  & v_i^j, 1 \leq j \leq n: & y_i^j > \dots > z_i^j
 & z_i^j > y_i^j > \dots > h \\
  & w_i^j, 1 \leq j < n: & z_i^j > x_i^j > \dots > y_i^{j+1} & x_i^j > \dots
  > y_i^{j+1} > z_i^j \\
  & w_i^n: & z_i^n > x_i^n > \dots > h & x_i^n > \dots > h > z_i^n
\\ \cline{3-4}
  V_3^p : &&
\multicolumn{2}{l}{ x_i^j > \dots > h
 \text{ for some } j \in \{\ell \condition
  e_{i} \in S_\ell\}}\\ 
\hline
\end{array}
\]
 
Every $x_i$ takes one last position and get his or her maximum partial score
with respect to~$c$.  For $e_i \in X'$, all $y_i^j$ take exactly one
first, one last, and a middle position in all remaining votes. For
$e_i \not \in X'$, all $y_i^j$ 
take middle positions only. So they
always get their maximum partial scores with respect to~$c$.  The
candidates $z_i^j$ also get their maximum partial scores with respect
to~$c$, since they always get one first position, one last position,
and a middle poisition in all remaining votes. Every candidate $x_i^j$
gets at most one first position and therefore does not exceed his or
her maximum partial score with respect to~$c$.  Since no candidate
exceeds his or her maximum partial score with respect to~$c$, 
candidate $c$ is a
winner in this extension of the list $V^p$ of partial votes.

Conversely, assume that $c$ is a possible winner for $(C,V)$.  Then no
candidate may get more points in $V^p$ than his or her maximum partial
score with respect to~$c$. Since at most $k$ different $x_i$ may take
a last position in~$V_1^p$, at least $n-k$ different $x_i$ must take a
last position in~$v_i$.
Fix any $i$ such that $x_i$ is ranked last in~$v_i$.
We now show that it is not possible that a candidate $x_i^j$ then
takes a first position in any
vote of~$V_3^p$.  Since $x_i$ takes the last
position in~$v_i$, $y_i^1$ takes a middle position in this vote and
gets one point.  The only vote in which the
score of $y_i^1$ is not
fixed is~$v_i^1$.  Without the points from this vote, $y_i^1$ already
gets $|V^p|-1$ points, so $y_i^1$ cannot get two points in~$v_i^1$,
and $z_i^1$ takes the first position in~$v_i^1$. Without the points
from $w_i^1$, $z_i^1$ gets $|V^p|$ points and must take the last
position in~$w_i^1$.  The first position in $w_i^1$ is then taken by
$x_i^1$, so $x_i^1$ cannot take a first position in any vote
from~$V_3^p$.  Candidate $y_i^2$ gets one point in~$w_i^1$, and by a
similar argument as above, $x_i^2$ is placed at the first position
in~$w_i^2$.
Repeating this argument,
we have that for each~$j$, $1 \leq j \leq n$, $x_i^j$ is placed at the
first position in $w_i^j$ and thus cannot take a first position in a
vote from~$V_3^p$.  This means that all first positions in the votes
of $V_3^p$ must be taken by those $x_i^j$ for which $x_i$ takes the last
position in a vote from~$V_1^p$.  This is possible only if the $x_i^j$
are not at the first position in~$w_i^j$.  Thus $z_i^j$ must take this
position.  Due to $z_i^j$'s maximum partial score with respect
to~$c$, this is possible only if $z_i^j$ takes the last
position in~$v_i^j$.  Then $y_i^j$ takes the first position in this
vote.  This is possible, since $y_i^j$ can take a middle position in
$v_i$ for $j=1$, and in $v_i^j$ for $2 \leq j \leq n$.  Hence all
$x_i^j$, where $x_i$ takes the last position in the votes of~$V_1^p$,
may take the first position in the votes of~$V_3^p$.  Thus, by the
definition of $V_3^p$ (which, recall, contains the vote
$T_j \succ C - \{T_j,h\} \succ h$ for each~$j$, $1 \leq j \leq n$,
where $T_j = \{x_i^j \condition e_i \in S_j\}$), the elements $e_i$
corresponding to those $x_i$ must form a hitting set of size at most
$k$ for~$\mathcal{S}$.~\end{proofs}

\section{Conclusions and Future Research}

In this paper, we have taken the final step to a full dichotomy
theorem for the {\sc Possible Winner} problem with unweighted votes
and an unbounded number of candidates in pure scoring rules.  Our
result complements the results of Betzler and
Dorn~\cite{bet-dor:j:towards-dichotomy} by showing that {\sc Possible
  Winner} is $\np$-complete for the pure scoring rule with vector
$(2,1,\dots,1,0)$, the one missing case
in~\cite{bet-dor:j:towards-dichotomy}.

Besides establishing this dichotomy theorem, our result has also other
consequences.  Since {\sc Possible Winner} is a special case of the
{\sc Swap Bribery} problem introduced by Elkind et
al.~\cite{elk-fal-sli:c:swap-bribery}, Theorem~\ref{thm:321-NP}
implies that this problem is $\np$-hard for the pure scoring rule with
vector $(2,1,\dots,1,0)$ as well.  Informally put, in a {\sc Swap
  Bribery} instance an external agent seeks to make a distinguished
candidate $c$ win the election by bribing some voters so as to swap
adjacent candidates in their preference orders (see
\cite{elk-fal-sli:c:swap-bribery} for formal details).

On the other hand, the {\sc Possible Winner} problem generalizes the
{\sc Coalitional Unweighted Manipulation} problem where a group of
strategic voters, knowing the preferences of the nonstrategic voters,
seeks to make their favorite candidate win by reporting insincere
preferences.  An instance of this manipulation problem can be seen as
a {\sc Possible Winner} instance in which all nonstrategic voters
report (sincere) complete linear orderings of all candidates, whereas
all strategic voters initially have empty preference lists, and the
question is whether they can extend them to complete linear orderings
of all candidates such that their favorite candidate wins.

The $\np$-hardness result of Theorem~\ref{thm:321-NP} has no direct
consequence for the complexity of this more special problem, and
neither so for other more special variants of {\sc Possible Winner},
such as {\sc Possible Winner with respect to the Addition of New
Candidates} (see Chevaleyre et
al.~\cite{che-lan-mau-mon:c:possible-winners-new-candidates-scoring},
Xia et
al.~\cite{xia-lan-mon:c:possible-winners-new-candidates-new-results},
and Baumeister et
al.~\cite{bau-roo-rot:c:two-variants-of-possible-winner}).  Note
that the complexity of the {\sc Coalitional Weighted Manipulation}
problem, where all votes are weighted and the weights of all
manipulators are known initially in addition to the weights and
preferences of the nonmanipulators, is well understood (see the work
of Conitzer et al.~\cite{con-san-lan:j:when-hard-to-manipulate}), and
even a dichotomy theorem for scoring rules due to Hemaspaandra and
Hemaspaandra~\cite{hem-hem:j:dichotomy-scoring} is known for weighted
votes.  However, the complexity of {\sc Coalitional Unweighted
  Manipulation} is still unknown for many voting systems, including
many scoring rules.  Only recently Betzler et
al.~\cite{bet-nie-woe:c:manipulation-borda} and Davies et
al.~\cite{dav-kat-nar-wal:c:complexity-and-algorithms-for-borda}
independently showed that {\sc Coalitional Unweighted Manipulation},
even for only two manipulators, is $\np$-complete for Borda elections,
where Borda with $m$ candidates is the scoring rule with vector $(m-1,
m-2, \ldots, 0)$.  Further complexity results regarding the
{\sc Coalitional Unweighted Manipulation} problem for various voting
systems are due to Faliszewski et
al.~\cite{fal-hem-sch:c:copeland-ties-matter,fal-hem-sch:c:copeland-manipulation},
Narodytska et
al.~\cite{nar-wal-xia:c:manipulation-of-nanson-and-baldwin}, Xia et
al.~\cite{xia-con-pro:c:scheduling-approach-to-coalitional-manipulation,xia-zuc-pro-con-ros:c:unweighted-coalitional-manipulation},
and Zuckerman et
al.~\cite{zuc-pro-ros:j:coalitional-manipulation,zuc-lev-ros:c:algorithm-for-coalitional-manipulation-under-maximin}.
None of these papers establishes a dichotomy theorem for manipulation
in the unweighted case, although dichotomy results for scoring rules
are now known for two of its generalizations, the {\sc Coalitional
  Weighted Manipulation} problem
(see~\cite{hem-hem:j:dichotomy-scoring}) and the (unweighted) 
{\sc Possible Winner} problem
(see~\cite{bet-dor:j:towards-dichotomy} and this paper).  For
future research, we propose to tackle the open problem of finding a
dichotomy result for {\sc Coalitional Unweighted Manipulation} in 
scoring rules.

\subsection*{Acknowledgments}

We thank the anonymous ECAI-2010 and IPL reviewers for their expert
comments on this paper that helped improving its presentation.

{\small
\bibliographystyle{elsarticle-num}
\bibliography{pw}
}

\end{document}